\documentstyle[12pt,epsf]{article}

\newcommand{\be}{\begin{equation}}
\newcommand{\ee}{\end{equation}}
\newcommand{\ba}{\begin{eqnarray}}
\newcommand{\ea}{\end{eqnarray}}
\newcommand{\nl}{\nonumber \\ & & \nonumber \\}

\newcommand{\parc}[2]{\frac{\partial #1}{\partial #2}}

\newcommand{\tr}{{\rm tr}\,}

\newcommand{\vs}{\vspace{0.5cm}}

\begin{document}
%
\vspace*{2.cm}
\begin{center}
{\Large Towards thermodynamical consistency \\ of quasiparticle
picture \\}
\vspace*{5mm}
{T.~S. Bir\'o, A.~A.~Shanenko\footnote{ Permanent address:
    Bogoliubov Laboratory of Theoretical Physics, Joint Institute for
    Nuclear Research, 141980 Dubna, Russia} and
V.~D.~Toneev$^{1}$ \\
%
\vspace*{3mm}
  Research Institute  for Particle and  Nuclear Physics, \\
 Hungarian Academy of Sciences, H-1525 Budapest, P.O.Box 49, Hungary \\ }

\end{center}
\vspace*{10mm}

\begin{abstract}
 \it The purpose of the present article is to call attention to some
realistic quasipa\-rti\-cle-based description of the quark/gluon
matter and its consistent implementation in thermodynamics.  A
simple and transparent representation of the thermodynamical
consistency conditions is given. This representation
allows one to review critically and systemize available phenomenological
approaches to the deconfinement problem with respect to their
thermodynamical consistency. A particular attention is paid to
the development of a method for treating the string screening in the
dense matter of unbound color charges. The proposed method yields
an integrable effective pair potential which can be incorporated
into the mean-field picture. The results of its application are in
reasonable agreement with lattice data on the QCD thermodynamics.
\end{abstract}

\vspace*{10mm}
{PACS numbers: 24.85.+p, 12.38.Aw, 12.38.Mh, 21.65.+f,
64.60.-i}

\newpage
\section{Introduction}

\vs With the advent of RHIC and LHC, there is a growing need for a
deeper understanding of various properties of the QCD matter at
high temperature and finite density.  At the moment, we are still
far from a satisfactory level in this respect, even for
equilibrium properties of quark--gluon plasma. Indeed, though such
a system can  in principle be approximated as a gas of quarks and
gluons, a fully perturbative calculation with these degrees of
freedom does in practice not work well at any reasonable
temperature since the perturbative series are badly  converged
due to infrared-sensitive contributions. On the other hand, the
QCD lattice calculation, the only systematic fully
nonperturbative method available, is restricted in the presence of
light dynamical quarks, and even more so in the presence of a
finite baryon density (see~\cite{karsch} where the current state
of art is summarized). Therefore, various phenomenological,
QCD-motivated models are called up for describing the
thermodynamics  of highly excited nuclear matter and its Equation
of State (EoS).

General arguments from QCD and lattice data tell that a kind of
string is developed between quarks and antiquarks at large
distance and it is natural to identify such $q\bar q$ system with
conventional mesons.  Treating quark and gluon propagation in the
confining QCD vacuum within non-Abelian SU(3) gauge theory, the
string dynamics was successfully applied to conventional mesons,
hybrids, glueballs and gluelamps. However, if such a string is
surrounded by unbound quarks and gluons, the $q\bar q$ system can
be excited not only in  color-singlet states, but also in color-octet
states or even dissociate into constituent elements. The latter
will signal, in general, on the deconfinement phase transition.
These phase transformations are intimately related to the change of string
properties (in particular, color charges  of quarks are screened
in quark-gluon environment):  string behavior becomes medium dependent.

By now, there is a number of simplified models for describing
static hadron properties as well as a highly excited, deconfined
state of quark matter,  the quark-gluon plasma (QGP). A common
feature of these models is that they all are based on a
quasiparticle picture, considering isolated particle-like degrees
of freedom and assuming that these quasiparticles are moving in a
background mean field. Two- and many-particle correlations are
included in the  mean-field contributions and in the modification
of the one-particle spectra. Well-known examples are the original
bag model and its later versions \cite{MIT,SLAC,CLOUDY},
phe\-no\-me\-no\-lo\-gi\-cal approaches with
tem\-pe\-ra\-ture-de\-pen\-dent bag con\-stant
\cite{levai+heinz,Kampfer}, string-motivated density-dependent
corrections to an ideal (massive or massless) quark matter
equation of state, the very consideration of hadrons  as
composite objects in QCD, excluded volume corrections
\cite{Hagedorn+Rafelski,Gorenstein}, and finally mixed phase
\cite{ton} and chemical mixture \cite{Knoll,BiLevZim} models
dealing with the transition between quark matter and hadron matter
in a phenomenological way.

The present paper concerns the quasiparticle description of the
QCD thermodynamics with the particular emphasis on the mean-field
treatment of in-medium strings. The paper is organized as
follows. In Sec.II we consider the thermodynamical consistency of
the quasiparticle description in general. Any phenomenological
approach, involving a quasiparticle interaction,  usually
operates with a Hamiltonian which may depend on thermodynamical
characteristics of the surrounding matter, like the temperature
$T$ and density $n$. As is known for a long time~(see, for
details, \cite{syu,gor}),  there exist certain restrictions to the
dependence of such a Hamiltonian on the thermodynamic variables.In
this section a transparent and useful representation of
these restrictions is derived
[see Eqs.(\ref{ctc1}) and (\ref{ctc1a})]  which directly
involves the quasiparticle spectra. The obtained representation of
thermodynamical consistency allows us to get an instructive
relation between a number of modern approaches, dealing with the
deconfinement problem in the framework of a quasiparticle picture,
as exemplified in the end of Sec.II. The QCD-motivated
interactions, in particular string-like interactions \cite{Satz},
are under investigation in Sec.III.  A comprehensive model of
string formation in the dense matter of unbound color charges is
developed, which supports the choice of the mean field proportional to an
inverse power of the color-charge density, as proposed in the
papers~\cite{ton,syu,olive,ropke,harz}. This quasiparticle scheme
is applied  in Sec.IV for  thermodynamics of the deconfined
QCD phase. The case when the system is in thermal equilibrium but
not in chemical one is also considered. The results are
summarized in the concluding Sec.V.

\section{Quasiparticle Hamiltonian}

\vs In order to obtain an ef\-fec\-tive qua\-si\-par\-tic\-le
des\-crip\-tion of a me\-dium made of un\-bound co\-lour charges,
one should operate with screened long-range potentials. A natural
way to introduce the screening in quark matter is based on using
the probability density $P(r)$ to form a string  of the length
$r$. It is worth noting that this investigating scheme has much
in common with another one which deals with the probability
density that the nearest neighbour occurs at a distance
$r$~\cite{ropke}. Both approaches involve thermodynamical
variables, which lead to a screened pair potential depending on
thermodynamical quantities. Due to such effects the quasiparticle
Hamiltonian becomes density and temperature dependent, which in
turn leads to a modification of a thermodynamical potential (e.g.
the Gibbs free energy). Eventually a nonideal EoS emerges.

\subsection{General structure of Hamiltonian}

We start with the general quasiparticle Hamiltonian
\be
\hat H = \sum_i\sum_k \epsilon_{ki}(T,n) \
             a^{\dag}_{{\bf k}i} \ a_{{\bf k}i}+ V \Phi(T,n).
\label{H} \ee
 Here $a^{\dag}_{{\bf k}i}$ and $a_{{\bf k}i}$ are
the usual creation and annihilation operators for quasiparticles
of the $i$-th sort with momentum $k$. They also may depend on
other internal degrees of freedom like spin, color, isospin, {\em
etc}. The volume $V$ is constant and large enough (infinite in
the thermodynamical limit). So in this limit, the summation over
quantum states labeled by $k$ can be replaced by a phase space
integral (including the internal degeneracy factor $d_i$):
\be
\sum_k \rightarrow V \, d_i \, \int \! \frac{d^3k}{(2\pi)^3}.
\ee
In general, the one-particle energy $\epsilon_{ki}(T,n)$ and the
background field contribution $\Phi(T,n)$ to the energy density,
both depend on the temperature $T$ and the set of particle
densities $n \equiv \{ n_1,n_2, \ldots \}$. Notice that the
quantity $V \Phi (T,n)$ is nothing else but the energy of the
quasiparticle vacuum. Generally speaking, it differs from the
vacuum of primordial particles, which leads to the $c$-number
term appearing in Eq.~(\ref{H}). In the present paper we consider
the situation when, similarly to the case of the Hartree-Fock
quasiparticles, the expectation value of the quasiparticle number
operator $\hat N_i = \sum_k  \, a^{\dag}_{{\bf k}i} a_{{\bf k}i}$
equals to the number of primordial particles $N_i=n_i V$. This
implies that we deal with the picture of quasiparticles
interacting and, thus, correlating with each other. In turn, the
expectation value of the Hamiltonian has to be equal to the mean
energy of the system under consideration. This leads to the
following relations:
\ba \label{h1} <\hat H > &=& E=V \sum_i
d_i\int \frac{d^3k}{(3\pi)^3}\;
                 \epsilon_{ki} \ \nu_{ki} + V\Phi,  \\
<\hat N_i > &=& N_i=V d_i \int \frac{d^3k}{(2\pi)^3}\; \nu_{ki},
\label{h2} \ea
with the occupation numbers $\nu_{ki}=\langle
a^{\dag}_{{\bf k}i}a_{{\bf k}i}\rangle$. There is another way of
calculating the mean energy $E=V\varepsilon(T,n)$ and mean
multiplicity $N_i=Vn_i$ which proceeds from a thermodynamical
potential rather than from Eqs.~(\ref{h1}) and (\ref{h2}). For
density- and temperature-dependent Hamiltonians these different
ways may lead to different results~(see, for
example~\cite{ton,syu}). Hence, in what concerns the dependence on
$n$ and $T$, the quasiparticle Hamiltonian will have a correct
structure only if {\it the thermodynamical consistency requirements}
(\ref{h1}) and (\ref{h2}) are satisfied when starting with either
the Hamiltonian or the thermodynamical potential.

\subsection{Chemical potentials}

Using temperature $T$ and number densities $n_i=N_i/V$ as basic
descriptive variables, the thermodynamical behavior and the appropriate
EoS can be derived from the corresponding thermodynamical potential,
the free energy $F(V,N,T)$:
\ba
F\equiv V f=T \sum_i \xi_i \ d_i \int\frac{d^3k}{(2\pi)^3}\,
             \ln\,\left(1-\xi_i \ e^{-z_{ki}}\right)
           +V\sum_i \mu_i \ n_i + V\Phi.
\label{F} \ea
Here $\xi_i=\pm 1$ is determined by the
quasiparticle statistics, $\mu_i$ stands for the chemical
potential of the quasiparticles of the $i$-th sort, and $z_{ki}$
is defined by
\be z_{ki}=\frac{\epsilon_{ki}-\mu_{ki}}{T}.
\label{zki} \ee
The simplest way of calculating the free energy
implies the use of the grand canonical ensemble when particle
numbers are known only in average and chemical potentials,
$\mu_i,$ are introduced instead of $N_i$ as descriptive
thermodynamical variables. To calculate the partition function in
this case, the quasiparticle Hamiltonian (\ref{H}) should be
modified to
\be \hat{H}' = \hat{H} - \sum_i\mu_i \hat{N}_i,
\label{H'} \ee
where $\hat{H}$ is defined by Eq.~(\ref{H}). We
recall that the physical meaning of $\mu_i$ is the energy loss by
removing a quasiparticle of the $i$-th species while the total
entropy and volume of the system are kept constant. This chemical
potential a priori has nothing to do with the fact whether this
particle really carries a conserved charge or not. However, there
are Lagrange multipliers associated to the conservation laws of
such charges like the baryon number, strangeness or electric
charge. In order to elucidate the difference between chemical
potentials in general and those associated to the conserved
charges, let us consider a particle mixture of many sorts whose
abundance is known only in average. The mixture components (not
necessarily all of them) carry some conserved charges. The conservation
of these charges is controlled by the appropriate chemical potential.
We denote such a charge of type $b$ carried by a
particle belonging to the $i$-th component of the mixture as
$q_{bi}$. Then for conserved quantities we have
\be Q_b = \sum_i
q_{bi} N_i. \label{Q} \ee
Usually there are more components than
the number of conserved charges. In particular, it is the case
for quark-gluon matter, to which we pay special interest in the
present paper. Besides the case of a one-component system, the
set of Eq.~(\ref{Q}) is insufficient for calculating all mean
numbers for the mixture components. Therefore, we need some
additional requirements which would allow us to determine the
particle numbers $N_i$ by making use of Eq.~(\ref{Q}).  The
chemical equilibrium is usually assumed and  then these additional
requirements are given by
\be \mu_i^{{\rm eq}} = \sum_b q_{bi}
\mu_b, \label{mui} \ee
where $\mu_b$ stands for the chemical
potential associated to the charge sort $b$. In the general case,
when the system is out of chemical equilibrium and the component
concentrations become time-dependent, the chemical potential can
be split into two parts,
\be \ \label{neqmu} \mu_i = \sum_b
q_{ib}\mu_b \, + \, \widetilde{\mu}_i\, , \ee
and Eq.~(\ref{H'}) is reduced to
\be \hat{H}' = \hat{H} - \sum_b \mu_b \hat{Q}_b -
\sum_i \widetilde{\mu}_i \hat{N}_i. \label{H'1} \ee
The quantities $\widetilde{\mu}_i$ describe the deviation from
chemical equilibrium in the thermally equilibrated system. They
are exactly zero in the chemical equilibrium limit, resulting in
the familiar relation
\be
 \hat{H}_{eq}' = \hat{H} - \sum_b \mu_b \hat{Q}_b.
\label{H'eq}
\ee
Below we shall investigate consistency of the quasiparticle
picture in thermodynamical treatment including the possibility
of deviations from chemical equilibrium in a mixture.

\subsection{Thermodynamical consistency}

As mentioned above, any approach starting with a thermodynamical
potential appears to be thermodynamically consistent. In other words,
if all the quantities of interest can be calculated only through the
derivatives of this thermodynamical potential, one is prevented from
encountering thermodynamical inconsistency. Problems arise, however,
when calculation can proceed not only from the constructed
thermodynamical potential but also from a more fundamental level,
some quasiparticle Hamiltonian at a given temperature and/or density.
In the last case the result may depend on the calculation method
unless the quantities $\epsilon_{ki}(T,n)$ and $\Phi(T,n)$ obey
relations derived in accordance with the consistency requirements
(\ref{h1}) and (\ref{h2}). These relations are below called as {\it
 conditions of  thermodynamical consistency}.

To elaborate on these conditions, let us consider a system with the
Hamiltonian defined by Eqs.~(\ref{H}) and (\ref{H'}). If the
abundance of all mixture components is known only in average, the
proper thermodynamical potential has the form
\be
\Omega = E - TS - \sum_i \mu_i \ N_i,
\label{Om}
\ee
with $S$ being the total entropy. The grand-canonical partition
function corresponding to $H'$ can readily be calculated
\be
Z\!=\!\tr \! (e^{-\hat H' /T})\! =\!
\left(\prod_{k,i} \sum_{n_{ki}}\!
\! e^{-z_{ki} n_{ki}} \right) \ e^{-\Phi V /T}=
e^{-\Omega/T}.
\label{Z}
\ee
Recall that in the Fermi-gas case the integer quantity $n_{ki}$
equals either $0$ or $1$, while in the Bose one it runs from $0$
to $\infty$. Hence, from Eq.~(\ref{Z}) it follows that
\be
\frac{\Omega}{V} = \,T \sum_i d_i \xi_i \int\frac{d^3k}{(2\pi)^3}
\ln \left(1-\xi_i e^{-z_{ki}}\right)+\Phi.
\label{Omega1}
\ee
Note that Eq.~(\ref{Omega1}) covers Eq.~(\ref{F}) with the definition
$F=E-T\,S$. It is well-known that for the Hamiltonian $H'$ the
average quasiparticle occupation number $\nu_{ki}$ is
\be
\nu_{ki}=\langle n_{ki} \rangle = \frac{1}{e^{z_{ki}}-\xi_i},
\ee
which together with Eq.~(\ref{Omega1}) leads to
\be
\frac{\Omega}{V} = \Phi\,-\,T\sum_i d_i \xi_i
  \int\frac{d^3k}{(2\pi)^3} \ \ln\left(1+\xi_i \ \nu_{ki}\right).
\label{Omega2}
\ee
The total differential of the thermodynamical potential $\Omega$
is given by
\ba
d\Omega &=& dV \left( \Phi - T \sum_i d_i \xi_i \int
   \frac{d^3k}{(2\pi)^3}  \ \ln
        \left(1+\xi_i \ \nu_{ki}\right) \right) \nl
 &-& V T \sum_i d_i \xi_i \int \frac{d^3k}{(2\pi)^3}
  \left(\frac{dT}{T}\ln\left(1+\xi_i \ \nu_{ki}\right)
   +\frac{\xi_i \ d\nu_{ki}}{1+\xi_i \
                                     \nu_{ki}}\right) + Vd\Phi.
\label{difomeg}
\ea
Since in the grand-canonical ensemble the quasiparticle densities
$n_i$ and average occupation numbers $\nu_{ki}$ are functions of the
temperature $T$ and chemical potentials $\mu_i$,  all the
differentials can be expanded in terms of $dT$, $dV$ and $d\mu_i$.
In particular, we have
\ba
d\nu_{ki}&=&- \nu_{ki} \ (1+\xi_i \ \nu_{ki}) \ d
\left( \frac{\epsilon_{ki}-\mu_i}{T}\right),\nl
d\epsilon_{ki}&=&\parc{\epsilon_{ki}}{\mu_j} \ d\mu_j
+ \parc{\epsilon_{ki}}{T} \ dT, \nl
d\Phi &=& \parc{\Phi}{\mu_j} \ d\mu_j + \parc{\Phi}{T} \ dT.\nonumber
\ea
Inserting this relations into Eq.~(\ref{difomeg}) and comparing then
the derived result with the general formula
$$
d\Omega = -p \ dV - S \ dT - \sum_i N_i \ d\mu_i\;,
$$
one can arrive at
\ba
\label{h1a}
\varepsilon\!&=&\!\sum_i d_i\!\int \!\frac{d^3k}{(2\pi)^3} \
      \epsilon_{ki} \ \nu_{ki}  + \Phi-\!\sum_i\mu_i C_i - TC_T \,,
\\
n_i\!&=&\! d_i\int \frac{d^3k}{(2\pi)^3} \ \nu_{ki} - C_i\;,
\label{h2a}
\ea
where
\ba
C_T &=& \parc{\Phi}{T}
      + \sum_j d_j\int \frac{d^3k}{(2\pi)^3}\;
                         \parc{\epsilon_{kj}}{T}\; \nu_{kj}, \nl
\quad C_i &=& \parc{\Phi}{\mu_i}
      + \sum_j d_j\int \frac{d^3k}{(2\pi)^3}\;
             \parc{\epsilon_{kj}}{\mu_i}\; \nu_{kj}.\nonumber
\ea
Equations (\ref{h1a}) and (\ref{h2a}) should be compared to the
consistency requirements given by Eqs.~(\ref{h1}) and (\ref{h2}).
It leads to the conditions $C_T=0$ and $C_i=0$.

This result can be rewritten in a more manageable form which allows
for relating our specific case based on Eq.~(\ref{H}) to a more general
one. To elucidate this connection let us consider the derivative matrix
$$
M_{ij} = \parc{n_i}{\mu_j}.
$$
Its elements can, in principle, have arbitrary values, and we expect
that the determinant of $M_{ij}$ is not zero. This is indeed the
case since the derivative matrix is given by the following implicit
relation:
$$
\parc{n_i}{\mu_j} = \frac{1}{T} \ d_i\int \frac{d^3k}{(2\pi)^3} \
\nu_{kj} \ (1+\nu_{kj})
\left( \delta_{ij} - \parc{\epsilon_i}{\mu_j} \right)
$$
leading to the matrix equation
$$
M_{jk} = A_{jk}  - \sum_i M_{ik} B_{ij}
$$
with
\ba
A_{jk}&=&\delta_{jk} \ \frac{1}{T} \ d_j\int \frac{d^3k}{(2\pi)^3}
\ \nu_{kj} \ (1+\nu_{kj}),
\nl
B_{ij}&=&\frac{1}{T} \ d_j\int \frac{d^3k}{(2\pi)^3} \ \nu_{kj} \
(1+\nu_{kj}) \   \parc{\epsilon_{kj}}{n_{ki}}.\nonumber
\ea
This matrix equation has a formal solution $M = (1+B)^{-1}A$. While
$B$ may have zero eigenvalues, $A$ does not, so the determinant $det
M = det A/det(1+B)$ cannot vanish. Using this information, the
equalities $C_T=0$ and $C_i=0$ can be rewritten as follows:
\ba
\label{ctc1}
\parc{\Phi}{T} + \sum_j d_j\int
\frac{d^3k}{(2\pi)^3}\;
\parc{\epsilon_{kj}}{T}\;\nu_{kj} = 0, \\
\parc{\Phi}{n_i}+
\sum_j d_j \int \frac{d^3k}{(2\pi)^3}\;
\parc{\epsilon_{kj}}{n_i}\; \nu_{kj} = 0~.
\label{ctc1a}
\ea
Equations (\ref{ctc1}) and (\ref{ctc1a}) represent a particular case of
the more general relations~\cite{ton,syu}
\be
\langle\parc{H_{eff}}{T}\rangle=0, \quad \quad \langle
\parc{H_{eff}}{n_i}\rangle=0.
\label{ctc2}
\ee
These conditions of the quasiparticle  consistency are reduced to
Eqs.~(\ref{ctc1}) and (\ref{ctc1a}) when the (temperature-  and
density-dependent) effective Hamiltonian $H_{eff}$ has the
quasiparticle form~(\ref{H}). Equations~(\ref{ctc2}) have first been
derived in~\cite{syu} (see also Ref.~\cite{ton}) under the
requirement that all the statistical ensembles of the system governed
by the density- and temperature-dependent Hamiltonian $H_{eff}$ yield
the same thermodynamics in the infinite volume limit. Thus, to avoid
thermodynamical inconsistency, the
constraints ~(\ref{ctc2}) should be satisfied while constructing the
effective quasiparticle Hamiltonians.

Let us emphasize that thermodynamical consistency is not sufficient
by itself when thermodynamics is constructed  starting from the level
of a thermodynamical potential but the Hamiltonian structure is ignored. In
this case nonphysical expressions can be involved even if there is no
problem with thermodynamical consistency and all the thermodynamical
quantities are derived by differentiating a thermodynamical
potential. We suggest that relations (\ref{ctc1}) and (\ref{ctc1a})
should also be employed in the situation like that to avoid
unreasonable expressions for quasiparticle spectra which can be met
in the literature. For instance, see the papers on the compressible
bag model~\cite{kagiyam}. It is thermodynamically consistent but
the quasiparticle spectra used there have nothing to do
with~(\ref{ctc1}) and (\ref{ctc1a}). Another example concerns the
approach of Ref.~\cite{risch} that has no problem with thermodynamics,
too. However, the shift of the chemical potential proposed in that
article is equivalent to the introduction of  a temperature-dependent
vector-type mean field. It is shown below  from Eqs.~(\ref{ctc1}) and
(\ref{ctc1a}) (see the next section, Example 1) that the mean field
like this can not depend on the temperature explicitly. By the way, it
is quite possible that the nonphysical feature of the quasiparticle
spectra used in the papers~\cite{risch} is an actual reason for
causality violation when the sound velocity is getting larger than the
velocity of light (for more details, see Ref.~\cite{prasad}).

Note that when temperature and density dependence of the effective
Hamiltonian is mediated only by some thermodynamical quantity
$\Lambda$, the consistency conditions~(\ref{ctc2}) are equivalent to
$\langle\partial H_{eff}/\partial \Lambda \rangle=0$. This relation
comes from the well-known expression $\delta F=\langle \delta H_{eff}
\rangle$~\cite{bog}, where $\delta F$ and $\delta H_{eff}$ stand for
infinitesimal changes of the free energy and Hamiltonian,
respectively. This expression can easily be derived by analogy with
the familiar Hellmann-Feynman theorem and, taken in conjunction with
the extremum condition for the free energy with respect to the
parameter $\Lambda$, leads to $ \langle\partial H_{eff}/\partial
\Lambda \rangle=0$.

\subsection{Quasiparticle spectra}

Conditions of thermodynamical consistency~(\ref{ctc1}) and
(\ref{ctc1a}) result in certain physical restrictions to the
mean-field potential depending on the structure of quasiparticle
spectra without coming into any detail of interaction between
constituents. We shall demonstrate that in a few cases used in
phenomenological treatments.\\

\noindent {\bf Example 1.} Let the energy of a quasiparticle of the
$i$-th sort moving with the 3-momentum ${\bf k}$ be approximated as
(for instance, see Refs.~\cite{ton,syu,olive,ropke,harz,heinz,kap,zim})
\be
\epsilon_{ki}(T,n)=\omega_i(k)+U_{i}(T,n),
\label{spectr1}
\ee
where $\omega_i(k)$ stands for the energy of the free particles of
the $i$-th sort and, as above, $n$  denotes the set of particle
number densities $ \{n_1,n_2,\ldots \}$. Generally, the mean-field
potential $U_{i}(T,n)$ is a function of temperature and particle
densities. The free particle energy can be given in either the
relativistic form, $\omega_i(k) = \sqrt{k^2 + m_{0i}^2}$, or in the
non-relativistic one, $\omega_i(k) =k^2/2m_{0i}$. It depends only on
the momentum ${ k}$ and bare particle mass $m_{0i}$. Then,
the conditions of quasiparticle consistency (\ref{ctc1}) and
(\ref{ctc1a}) are reduced to the following equations:
\be
\parc{\Phi}{T} + \sum_j n_j \ \parc{U_j}{T} = 0, \quad \quad
\parc{\Phi}{n_i}+ \sum_j n_j \ \parc{U_{j}}{n_i} = 0.
\label{ctc3}
\ee
As it is seen, ignoring the background field contribution $\Phi(T,n)$,
like, for example, in the papers~\cite{harz,heinz}, results in a loss
of thermodynamical consistency. It is important to note that
Eqs.~(\ref{ctc3}) are only then compatible, if the mean field $U_{i}$
does not depend explicitly on the temperature. Indeed, the first
equation in~(\ref{ctc3}) has an integral of the form $$ \sum_{j}\;
n_{j}U_{j}+\Phi=\varphi,$$ where $\varphi=\varphi(n)$ is an arbitrary
function of the quasiparticle densities. By differentiating this
expression with respect to $n_i$, we get
$$
\parc{\varphi}{n_i}=U_{i}+\sum_{j} n_{j} \ \parc{U_j}{n_i}+
\parc{\Phi}{n_i}.
$$
Taken in conjunction with the second equation in ~(\ref{ctc3}), the
obtained relation is reduced to $\partial \varphi/\partial n_i=U_i$.
It follows then that $U_i$ and $\Phi$ are temperature-independent
functions. In other words, when quasiparticle spectra are defined by
Eq.~(\ref{spectr1}), the thermodynamically consistent mean-field
potential may depend only on particle densities: $\;U_i=U_i(n), \;
\Phi=\Phi(n)$. Note that this important point is missed in some
papers~\cite{anchish}, where the excluded volume effects are treated
by means of the mean--field approximation.

Equations~(\ref{ctc3}) lead to one more interesting result
\be
\parc{U_i}{n_j}=\parc{U_j}{n_i}.
\label{krest}
\ee
This crossing relation, first presented in~\cite{zim}, follows from
the second equality in Eqs.~(\ref{ctc3}). To derive
Eq.~(\ref{krest}), one should differentiate the second equation
in~(\ref{ctc3}) with respect to $n_l$. Then, by interchanging the
indices $l$ and $i$ and comparing the obtained expression with the
previous one, we arrive at Eq.~(\ref{krest}). By doing so one should
keep in mind that
$$ \frac{\partial^2 U_j}{\partial n_l \ \partial n_i}=
\frac{\partial^2 U_j}{\partial n_i \ \partial n_l}\, ,\quad \quad
      \frac{\partial^2  \Phi}{\partial n_l \ \partial n_i}=
      \frac{\partial^2 \Phi}{\partial n_i \ \partial n_l}. $$
This is valid, provided the second derivatives of $U_j$ and $\Phi$
are continuous functions of $T$ and $n$, which is usually the case.
Note that Eq.~(\ref{krest}) is very useful when dealing with the
mean fields $U_i$ for many-component systems. For example, see the
investigation of quark--hadron interactions in \cite{ton}. The
crossing relation (\ref{krest}) should be kept  under constructing the
mean fields acting on quasiparticles of different species in a
many-component system. Otherwise, the thermodynamical
consistency can be lost.  As an example, one can point out Ref.~\cite{kap},
where mean fields were chosen as $U_i\propto n_{tot} \ (\widetilde
m/m_{0i})^{\delta}$~with $\delta=1$ or $2$. Here $\widetilde m$ denotes the
nucleon mass and $n_{tot}=\sum n_j$.\\

\noindent
{\bf Example 2.} Another popular form of quasiparticle spectra
$\epsilon_{ki}(T,n)$ is given as
\be
\epsilon_{ki}(T,n)=\sqrt{k^2+m^2_{i}(T,n)},
\label{spectr2}
\ee
{\em i.e.} an effective quasiparticle mass $m_i \equiv m_{i}(T,n)$
is introduced in a way similar to scalar mass in the  relativistic
mean-field theory~\cite{walecka}. In this case Eqs.~(\ref{ctc1}) and
(\ref{ctc1a}) can be rewritten as follows:
\be
\parc{\Phi}{T} + \sum_j n^{(s)}_j \ \parc{m_j}{T} = 0, \quad \quad
\ \parc{\Phi}{n_i}+ \sum_j n^{(s)}_j  \
\parc{m_{j}}{n_i} = 0,
\label{ctc4}
\ee
where the Lorentz-scalar quasiparticle density $n^{(s)}_{i}$ is
defined by
\be
n^{(s)}_{j}=d_j\int\frac{d^3k}{(2\pi)^3} \ \nu_{kj}\;
\frac{m_j}{\sqrt{k^2+m_j^2}}\; .
\label{scalden}
\ee
For the sake of simplicity, let us limit ourselves to the case of
one sort of quasiparticles. Differentiating the first equation in
(\ref{ctc4}) with respect to $n$ and the second one with respect to
$T$, we arrive at
\be
\parc{n^{(s)}}{T}\;\parc{m}{n}-
\parc{n^{(s)}}{n}\;\parc{m}{T}=0,
\label{grad}
\ee
provided the mixed second derivatives of $\Phi$ and $m$ are equal to
each other.  As follows  from Eq.~(\ref{grad}), the gradients of
functions $n^{(s)}(T,n)$ and $m(T,n)$ are parallel vectors in the
$(T,n)$ plane. Hence, $m$ is left constant along any line where
$n^{(s)}$ is constant. Since $n^{(s)}(T,n)$ and $m(T,n)$ are
differentiable functions of $T$ and $n$,  the $(T,n)$ plane is
densely covered by lines of constant $n^{(s)}$. Therefore, in a
thermodynamically consistent model with one sort of quasiparticles,
whose energy is defined by~(\ref{spectr2}), the effective
quasiparticle mass depends on temperature and quasiparticle density
{\em only through the scalar density.} The Walecka model~\cite{walecka}
without vector field (for zero baryon density) is a particular case
of the considered variant. Another example can be found in the paper
of Boal, Schachter and Woloshin in Ref.~\cite{olive}, where
interactions in the quark-gluon plasma are described by introducing
the effective masses of quarks and gluons depending on sum of the
color-charge densities. As it has been proven above,  this version is
inconsistent. One can expect that  the case of many
quasiparticle species  is described by equations  similar to Eq.~(\ref{ctc4}),
{\em  i.e.} $m_i$ is a function of the
set of $n_i^{(s)}$. As a consequence $\Phi$ depends on $T$ and $n$
through the scalar densities too and satisfies
\be
\Phi = - \int \, \sum_j
n_j^{(s)} \, dm_j.
\ee
Deconfinement models dealing with the tem\-pe\-ra\-ture- and
den\-si\-ty-de\-pen\-dent mas\-ses of quarks and
glu\-ons~\cite{levai+heinz,golov,rischke} are also related to the
Example 2. Here it is often assumed $n^{(s)} \propto m^3$ and $\Phi \propto
m^4$. A purely tem\-pe\-rature-de\-pen\-dent bag con\-stant,
$\Phi(T)$, without mass modifications on the other hand is
inconsistent. The same is related to the situation when
temperature-dependent masses without the background term are used
\cite{pin}. To go in more detail, see also the papers~\cite{gor}.\\

\noindent {\bf Example 3.} If the mean field in Eq.~(\ref{spectr1})
is scaled with  some coupling constant, $U_i=g_i\,U$~(see, for
example, Refs.~\cite{ton,syu,zim}), then conditions of thermodynamical
consistency take the form
\be
\parc{\Phi}{T} + \rho \parc{U}{T} = 0, \quad \quad
\parc{\Phi}{n_i}+  \rho \ \parc{U}{n_i} = 0
\label{ctc5}
\ee
with
$$\rho=\sum_j g_j \ n_j.$$
Using the procedure similar to that
described in the previous example, one can be convinced that the
density dependence of the mean field $U$ is mediated by $\rho(n)$ only.
If $g_i$ is proportional to the baryon number $b_i$ of the
quasiparticle $i$: $g_i=g b_i$, then we get $\rho=g\sum_i b_i n_i\equiv
g n_b$, where $n_b$ is the total baryon density. Similar situation is
realized when the quasiparticle interaction is mediated by a vector
field.

Now let us return to Ref.~\cite{kap} mentioned at the end of the
Example 1. Eqs.~(\ref{ctc5}) suggest how one can correct the
mean field $U_i=n_{tot} \ (\widetilde m/m_{0i})^{\delta}$, used in
this paper, in such a way that to keep the relation $U_i \propto
(\widetilde m/m_{0i})^{\delta}$. It turned out that the unique
solution is given by $U_i=\varphi(\rho)\ (\widetilde m/m_{0i})^{\delta}$,
where instead of $n_{tot}$ we use $\rho=\sum (\widetilde m/
m_{0j})^{\delta} n_j$ and an arbitrary function $\varphi(\rho)$ which
can be chosen as $\varphi(\rho)=\rho$.

Sometimes it is convenient to subdivide the full set of coupling
constants into the two groups: $g^{(a)}_{i}<0$ (corresponding to the
attractive interaction) and $g^{(r)}_{i}>0$  (related to the repulsive
interaction,~see, for example, Ref.~\cite{zim}). In this case
$U_i=g^{(r )}_{i} U_r-|g^{(a)}_i| U_a$, and we can expect that the
repulsive-interaction component $U_r$ is a function of $\rho_r=
\sum_i g^{(r)}_i n_i$, while the attractive one $U_a$ depends on
quasiparticle densities through $\rho_a=\sum_i g^{(a)}_i n_i$.\\

\noindent
{\bf Example 4.} By analogy with the approximation $U_i=g_i U$
considered in the Example 3, the effective quasiparticle mass of
Example 2 can also be scaled as $m_i=g_i \ M(T,n)$. In particular,
$M(T,n)$ can be the constituent quark mass, whereas $g_i$ is the
number of quarks in the baryon cluster of the $i$-th sort. Generally, a
 cluster of $g_i$ constituents may consist of quarks,
antiquarks and gluons ({\em i.e.} forming baryons, mesons, hybrids,
glueballs), as well. By doing so, we get
\be
\parc{\Phi}{T} +
\rho^{(s)} \ \parc{M}{T} = 0, \quad  \quad
\parc{\Phi}{n_i}+\rho^{(s)} \ \parc{M}{n_i} = 0
\label{ctc6}
\ee
with $$ \rho^{(s)}=\sum_j g_j \ n^{(s)}_j.  $$ The thermodynamics of
such a system depends on the descriptive variables $T$ and $n$
through the quantity $\rho^{(s)}(T,n)$.

\section{Mean-field treatment of string interactions}

Until now we have discussed general restrictions to the phenomenological
Hamiltonian due to the thermodynamical consistency. It can be further
elaborated by specifying the interaction between generic
constituents. The quasiparticle properties ought to be derived from
this underlying interaction. In our phenomenological treatment we
consider a particular case of strong pair interaction mediated by
strings stretched between color charges. The main difficulty here is
that such a system is plagued with long-range interaction, and only
in-medium screening renders the problem treatable,  even in the
weak-correlation approximation.

Strings are particular QCD field-constructions involved in the
interaction between two color charges in vacuum. We consider here how
this interaction will behave in a medium consisting of point-like
color charges. Let some reference color charge creating a string be
placed at the origin of coordinates and other color charges be
distributed around with the density $n(\ell)$ where $\ell$ is the
distance from the reference charge. Physically, we can expect that
the string formation is characterized by a probability depending on
its length, but not on its formation history. Let $P(\ell)d\ell$ be
the probability for a string to have a length between $\ell$ and
$\ell+d\ell$. The quantity $P(\ell)$ can be represented as a product
of two factors:
\be
P(\ell) = \left( 1-\int_0^{\ell} dx\;P(x)\right)\,w(\ell)\,.
\label{6}
\ee
The first factor in Eq.~(\ref{6}) is the probability
for the string to have the
length not less than $\ell$. The second factor $w(\ell)$ is related
to the conditional probability, $w(\ell)d\ell$,  meaning that a
string is formed between $\ell$ and $\ell+d\ell$ (provided it has
already reached the length $\ell$).

For gradually growing strings the quantity $w(\ell)$ can be obtained
invoking the arguments similar to those used in calculation of the
mean-free path of a particle moving through a medium in a given
direction. In this scenario we assume that a string is caught by any
color charge within a cylinder of the radius $a$ and with axis along
the considered string direction. This leads to
\be
w(\ell)=\pi a^2 \ n(\ell).
\label{rigid}
\ee
The factor  $\pi a^2$ is interpreted as string cross section
accounting also for a lack of  string dynamics. The effective radius,
$a$, may depend on the medium.

In another scenario the strings are assumed to wildly fluctuate in
direction. The gross factor $w(\ell)$ is rather well approximated
in this case by the relation
\be w(\ell) = 4\pi \ \ell^2 \ n(\ell).
\label{fluct}
\ee
Eq.~(\ref{fluct}) having an additional $\ell^2$ factor can be derived
by analogy to Eq.~(\ref{rigid}) if the whole area of the spherical
shell, $4\pi\ell^2d\ell$, at the distance $\ell$ is taken into
account. Here all possible partner charges, located at the distance
$\ell$, potentially participate in the screening. Note that in both
cases $n(\ell)$ stands for the number density of potential partners,
on which a string of length $\ell$ can be closed.

Usually it turns out to be more convenient to deal with the
integro--differential equations rather than with the integral ones.
After differentiating Eq.~(\ref{6}), we arrive at
\be
\frac{dP}{d\ell} = - w(\ell) \ P(\ell) + \frac{dw}{d\ell} \
\left( 1 - \int_0^{\ell} P(x) \ dx \right).
\ee
After substituting here the integral definition (\ref{6}) we get
\be
\frac{dP}{d\ell} = - w(\ell) \ P(\ell) + \frac{dw}{d\ell}
\ \frac{P}{w}.
\ee
This ordinary differential equation is separable and has
the following explicit solution
\be
P(\ell) \, = b\, w(\ell)\,e^{- \int_0^{\ell}w(x) \ dx}.
\label{scr}
\ee
The integration constant $b$ is determined by
normalization of the probability density:
\be
\int_0^{+\infty} \, P(x)\ dx \, = \, 1.
\ee
Let us consider now Eqs.~(\ref{rigid}), (\ref{fluct}) and
(\ref{scr}) in more detail. Neglecting the spatial charge
correlations ({\em i.e.} taking $n(\ell) = n = const$),
Eqs.~(\ref{rigid}) and (\ref{scr}) (belonging to the first,
straight string scenario) results in an exponential screening
\be
P(\ell) = \pi a^2 n e^{ - \pi a^2 n \ \ell}. \label{scr1}
\ee
>From here, the probability for a string to be shorter than $\eta$
is given by
\be Q(\eta) = \int_0^{\eta} P(x) \ dx = 1 - e^{ - \pi a^2 n \
\eta}
\label{scr1a}
\ee
and the average string length is
\be \langle \ell \rangle \, = \,
\int_0^{\infty} x \, P(x) \, dx = \frac{1}{\pi a^2 \, n}.
\label{scr1b}
\ee
Noting that the probability distribution (\ref{scr1}) can be
represented in the form
\be P(l) \propto
e^{-E(\ell )/E(  \langle \ell \, \rangle)} = e^{- \ell/ \langle
\ell \, \rangle}~,
\label{scr1c}
\ee
one sees that short strings are energetically favored. This looks
as a natural conclusion, but nevertheless it is not trivial,
because ``energy'' arguments have not been involved explicitly in
 reasoning.

Within the second scenario of the string screening (i.e. strings are
wildly fluctuating, see Eq.~(\ref{fluct})), the cor\-res\-pon\-ding
probability density becomes
\be
P(\ell) = 4\pi \,n \, \ell^2 e^{-4\pi \, n \, \ell^3/3}.
\label{scr2}
\ee
This expression covers the result of the papers~\cite{ropke} where
the probability of string formation has been calculated in the
nearest--neighbor approximation. The probability of the string length
to be less than $\eta$ has now the form
\be
Q(\eta) = \int_0^{\eta} P(x) \, dx = 1 - e^{ -\frac{4}{3}
\pi  \, n \, \eta^3},
\label{scr2a}
\ee
and the average string length is given by
\be
\langle \ell \rangle \, = \,
\Gamma(\frac{1}{3})\left( \frac{3}{4\pi\,n} \right)^{1/3}
\label{scr2b}
\ee
with $\Gamma(\ldots)$ being  Euler's Gamma function. By analogy with
the representation (\ref{scr1c}), Eq.~(\ref{scr2}) can be rewritten
as
\be
P(\ell) \propto \left( \ell \right)^2 \exp {
\left( - \left( \Gamma(\frac{1}{3})
\frac{ \ell }{ \langle \ell \rangle}  \right)^3 \right) },
\label{scr2c}
\ee
which also agrees with the argument that short strings are favorable.

As noted above  the straight-string scenario based on
Eq.~(\ref{rigid}) involves the effective string radius, $a$, which
may depend on thermodynamic variables. Indeed,  the string survives
only in the case when the characteristic length $a$ does not exceed
the mean distance between neighboring color charges, $r_0 =(4\pi n
/3)^{-1/3}$. Therefore, operating with Eq.~(\ref{rigid}) and
Eqs.~(\ref{scr1})-(\ref{scr1c}), we should employ $a \leq r_0$. Now,
estimating $a \approx c r_0$~($c < 1$ is some constant), we obtain
from Eq.~(\ref{scr1b})
\be
\langle \ell \rangle \, = \,\left(\frac{4}{3 c^3}\right)^{2/3} \
\frac{1}{(\pi n)^{1/3}}
\label{scr1bb}
\ee
in accordance with Eq.~(\ref{scr2b}) under the choice $c = 2/
\sqrt{3\Gamma(1/3)} \approx 0.7 < 1$. Summarizing, the thermodynamics
of string interactions is ruled by the average length of in-medium
strings. Thus, in spite of differences in $P(\ell)$, both considered
scenarios of the string screening lead to qualitatively similar
thermodynamical pictures. It is interesting that the
density-dependent interpretation of $a$ in Eq.~(\ref{scr1}) leads to
a density-dependent screening:
$$
e^{-\pi a^2 n r}=e^{-M_{\rm scr} r}
$$
with the screening mass $M_{\rm scr} =(3c^3/4)^{2/3} (\pi n)^{1/3}$.
At sufficiently high temperatures the QCD thermodynamics approaches
the Stefan--Boltzmann regime where $n \sim T^{3}$. This yields
$M_{\rm scr} \propto T$, what is nicely consistent with perturbative
QCD.

The modification of the energy density due to presence of in-medium
strings can be constructed as
\be
\Delta \varepsilon =  n  \ \sigma \ \langle \ell \rangle
\label{enmod}
\ee
with the string tension $\sigma$. In this case the color constituents
of the system are affected by the following mean field:
\be U=\sigma  \
\langle \ell \rangle=A \ n^{-\gamma},
\label{meanf}
\ee
where the constants $A$ and $\gamma$ carry information about the sort
of color charges and character of the in--medium string screening.
Note that the results of lattice simulations for $SU(3)$ symmetry can
be approximated by $\gamma \approx 2/3$~\cite{ton,syu}. This is in
qualitative agreement with our rough estimate $\gamma \sim 1/3$
neglecting the spatial correlations of color charges. Indeed, in
respect to the thermodynamical character of the EoS, the only fact
is decisive, that $\gamma$ lies between $0$ and $1$.

Concluding this section, we sketch how one should, in principle, deal
with the case when  spatial correlations of color charges are taken
into account. Let us consider a reference color charge placed at the
origin. The important point is that the ratio
\be
g(\ell)=\frac{n(\ell)}{n}
\label{radial}
\ee
is nothing else but the radial distribution function~\cite{shan}
which determines the pair particle correlations in the {\it uniform}
system of color charges with the density $n$ given by
$$
n=\lim_{\ell \to \infty}n(\ell).
$$
As follows from Eq.~(\ref{radial}), Eqs.~(\ref{scr1}) and
(\ref{scr2}) operate with $g(l)=1$, which corresponds to neglecting
the spatial correlations of color charges. To go beyond this
simplification, one should replace $n(\ell)$ by $n g(\ell)$ in
Eq.~(\ref{scr}). In particular, using Eq.~(\ref{rigid}), one can
derive the following equation:
\be
P(\ell)= b\,\pi a^2 \, n  \, g(\ell)  \
e^{ - \pi a^2 \, n \int_0^{\ell} g(x) \, dx}~,
\label{radial1}
\ee
the average length of the in--medium strings being dependent on the
charge--charge spatial correlations. Note that for $\ell \to \infty$
the quantity $g(\ell)$ tends to $1$, and we arrive at the exponential
decay of $P(\ell)$ with the screening factor which is  the same as in
Eq.~(\ref{scr1}), but with the different normalization constant $b$.
In principle, this difference can lead to another estimate of
$\gamma$ being closer to the lattice result mentioned above. Thus,
the probability density $P(\ell)$ should not be identified with the
radial distribution function as it has been done in the
paper~\cite{ropke}.

\section{Application to QCD thermodynamics}

The developed technique allows one to construct a thermodynamical
potential in a self-consistent way starting from the microscopic
level. The Hamiltonian structure of a particular phenomenological
model is defined by the physical assumptions used. Some simple
models for the QCD thermodynamics are considered below in order to illustrate
convenience and power of the conditions of thermodynamical consistency
(\ref{ctc1}), (\ref{ctc1a}) and to show the rationality of our treatment
of in-medium strings.

\subsection{Ideal gas in a bag-like model}

Let us consider the ideal gas of particles whose one-particle
spectrum is independent of medium parameters,
\be
\epsilon_{ki}(n,T) = \omega_i (k) \ .
\ee
Then Eqs. (\ref{ctc1}) and (\ref{ctc1a}) have the form
\be \frac{\partial \Phi}{\partial T}= 0, \quad  \quad
\frac{\partial \Phi}{\partial n_i}= 0
\ee
leading to a constant background energy $\Phi=B$. It is frequently
associated with the bag constant. In this relatively simple case
the chemical potentials are nevertheless determined by the set of
the implicit equations
\be n_i= d_i \,\int \! \frac{d^3k}{(2\pi)^3}\
\frac{1}{e^{(\omega_i(k)-\mu_i)/T}-\xi_i}.
\label{IDEAL-GAS}
\ee
However, in the classical approximation we have $\nu_{ki} << 1$
and Eq.~(\ref{IDEAL-GAS}) is reduced to the expression
$$
n_i = d_i \, \int \! \frac{d^3k}{(2\pi)^3} \
e^{-(\omega_i(k)-\mu_i)/T} = \chi_i(T) \ e^{\mu_i/T}.
$$
Here the second equality defines $\chi_i(T)$
which relates to the chemical potential as
\be
\mu_i= T \,\ln \left(\frac{n_i}{\chi_i(T)}\right).
\label{bmu}
\ee
Thus, for the internal energy density we obtain
\be
\varepsilon = 3T \sum_i n_i \, + \, B,
\ee
whereas the pressure is given by
\be
p= T \sum_i n_i-B.
\ee
These equations constitute the classical approximation to
the familiar MIT bag
model~\cite{MIT,cleymans} being a popular approach of investigating
the thermodynamics of the quark-gluon plasma.

\subsection{Temperature-dependent scalar mean field}

The temperature-dependent scalar mean field accounts for a
temperature-dependent mass. In this situation for the one-component
system the quasiparticle spectrum is given by (cf. (\ref{spectr2}))
\be
\epsilon_k(T,n) = \sqrt {k^2 + m^2(T)}~, \label{meff}
\ee
the
corresponding scalar density (cf. Eq.~(\ref{scalden})) depending
only on temperature. The gluon and quark plasma (at zero baryon density)
is of particular interest where approximately \be m^2(T) =
m^2_0 +g^2 T^2~. \ee Neglecting the derivative of the slowly
changing temperature function $g(T)$, from  Eq.(\ref{meff}) we
obtain
\be \frac{\partial \epsilon_k}{\partial T} =
\frac{m(T)}{\epsilon_k}\ m'(T) = \frac{g^2T}{\epsilon_k}.
\ee
Hence, the $C_T=0$ constraint leads to
\be \frac{\partial
\Phi}{\partial T} + g^2 T d \int \frac{d^3k}{(2\pi)^3} \
\frac{\nu_k} {\epsilon_k} = 0~.
\ee
One obtains for the equilibrium state
($\mu=0$):
\be
\varepsilon={\cal K} T^4 + B\,,\quad p=\frac{1}{3}
                                    {\cal K} T^4 -B,
\ee
where ${\cal K}$ is given by the integral
\be
{\cal K}=d\int \frac{d^3x}{(2\pi)^3}\;
\frac{1}{e^{\sqrt{x^2+g^2}}-\xi} \,
\frac{x^2+\frac{3}{4}g^2}{\sqrt{x^2+g^2}}.
\ee
The $g=0$ case provides the original MIT-bag EoS.

\subsection{Density-dependent mean field}

Our next example deals with a system of quasiparticles of a single
sort with the density-dependent spectrum justified in Sec. 3:
\be
\label{sp1} \epsilon_k(T,n) = \omega (k) + A \ n^{-\gamma}\,,
\ee
where $A$ and $\gamma$ have been discussed above. In this case
Eqs.~(\ref{ctc1}) and (\ref{ctc1a}) give rise to the following
form of the background energy density:
\be \label{sp2} \Phi
(n)=\frac{\gamma}{1-\gamma} A \ n^{1-\gamma}\;.
\ee
Hence, in the Boltzmann approximation we find
\be
\mu=T \ \ln \frac{n}{\chi(T)} +  A \ n^{-\gamma}\,,
\ee
whereas for the internal energy and pressure one can derive
\ba
\varepsilon&=& \frac{n d}{\chi(T)}
\int\frac{d^3k}{(2\pi)^3}
\;\omega(k)\;e^{-\omega(k)/T}+\frac{1}{1-\gamma}\  A \ n^{1-\gamma},\\
p&=&n T-\frac{\gamma}{1-\gamma} \ A \ n^{1-\gamma}.
\label{pdendep}\ea
As follows from our consideration in Sec.3, $0<\gamma<1$. In this
case the chemical potential grows with decreasing density ({\em e.g.}
due to string pulling) and therefore  such systems reveal a strong
tendency to form clusters. Free sources of strings, or long strings
respectively, will eventually be purged out of the system.

\subsection{Massless gluons with string interaction}

Let us consider in more detail the case given by
Eq.~(\ref{sp1}) for massless $SU(3)$ gluons. The
choice $\gamma = 1/3$ and $A = (2/3)\sigma$ satisfies
Eqs.~(\ref{scr1bb}) and (\ref{meanf}) at $a \approx r_0=(3/4\pi
n)^{1/3}$. The relation between density and {\em
non-equilibrium} chemical potential (see Eq.~(\ref{neqmu})) is now
given by
\be n=d\int \frac{d^3k}{(2\pi)^3}\;
   \frac{1}{e^{(k+\frac{2}{3}\sigma n^{-1/3}-\mu)/T}-1} \ .
\ee
In the Boltzmann approximation, which is quite appropriate to the
system of interest, this equation yields
\be
\mu = \frac{2}{3}\sigma n^{-1/3} + T \  \ln \
\frac{n}{\chi(T)},
\ee
where $\chi(T)$ is proportional to $T^3$:
\be
\chi(T) = (\lambda T)^3,\quad \lambda =\left(
                                \frac{d}{\pi^2}\right)^{1/3}.
\ee
We seek for the solution of the chemically equilibrium state
defined by $\mu=0$,
\be
n_{eq} = \chi(T) \ {\displaystyle e^{ -\frac{2}{3}\sigma
n_{eq}^{-1/3}/T}}~.
\ee
This can be transformed into a simple transcendental equation
by denoting
\be
z=T \ n^{-1/3}_{eq}, \quad \quad \widetilde{\sigma} =
\frac{2}{9}\ \frac{\sigma}{T^2} \ .
\ee
We get
\be
\lambda z=e^{\widetilde{\sigma}z} \ .
\label{chem}
\ee
This equation has no real solution above the value of
$\widetilde{\sigma}$ corresponding to a temperature $T_{\rm chem}$.
At this chemical critical temperature the l.h.s. and r.h.s. of
Eq.~(\ref{chem}) and their derivatives should be equal to each
other. So, the last condition gives:
\be
\lambda=\widetilde \sigma \ e^{\widetilde \sigma z_{\rm cr}}~.
\ee
Comparing with Eq.~(\ref{chem}), we obtain
\be z_{\rm cr} =
1/\widetilde{\sigma}, \quad \quad  \lambda=\widetilde{\sigma} e \ ,
\ee
where $e=e^1$. Finally re-expressing the temperature, we arrive at
the result that below the value
\be T_{\rm chem}=\sqrt{\frac{2e}{9\lambda}} \ \sqrt{ \sigma }
\label{tchem}\ee
there is no equilibrium solution for the string-like EoS. Assuming
$SU(3)$ symmetry for massless gluons with $d=16$ degrees of
freedom, we have $\lambda\approx1.175$ and arrive at the
estimate
\be
T_{{\rm chem}} \approx 0.718 \ \sqrt{\sigma} \ , \label{tchem1}
\ee
which for  $\sigma = 0.22 \ GeV^2$ gives $ T_{\rm chem}\approx 337$
MeV. It is noteworthy that a similar relation  between the color
deconfinement temperature and string tension, $T_c = (0.6-0.65)
\sqrt{\sigma} = 280-305$ MeV, has been obtained in the Monte
Carlo simulation of the lattice $SU(3)$ quenched
QCD~\cite{karsch,gboyd}.

\begin{figure}
\begin{center}
\leavevmode
\epsfxsize=100mm
\epsfbox{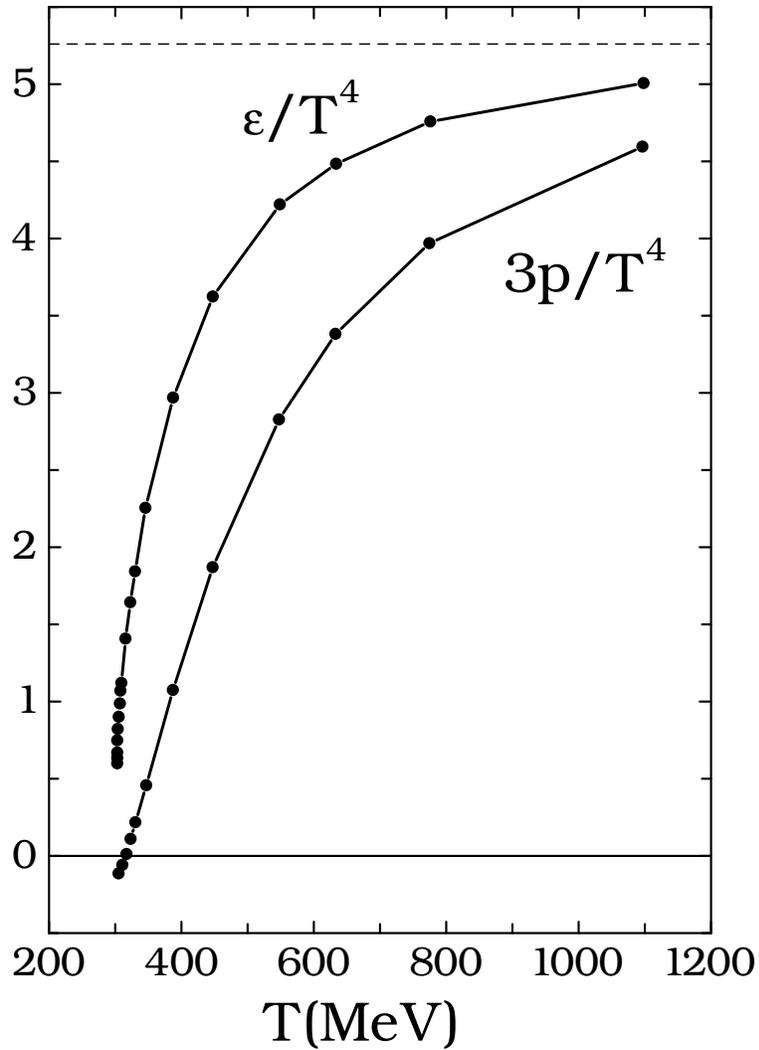}
\end{center}
\caption{Normalized energy density and pressure of a massless
gluon gas with string-like interaction. Dashed line demonstrates
the Stephan-Boltzmann regime corresponding to the case of a gas
of noninteracting gluons.}
\label{fig1}
\end{figure}

\begin{figure}
\begin{center}
\leavevmode
\epsfxsize=100mm
\epsfbox{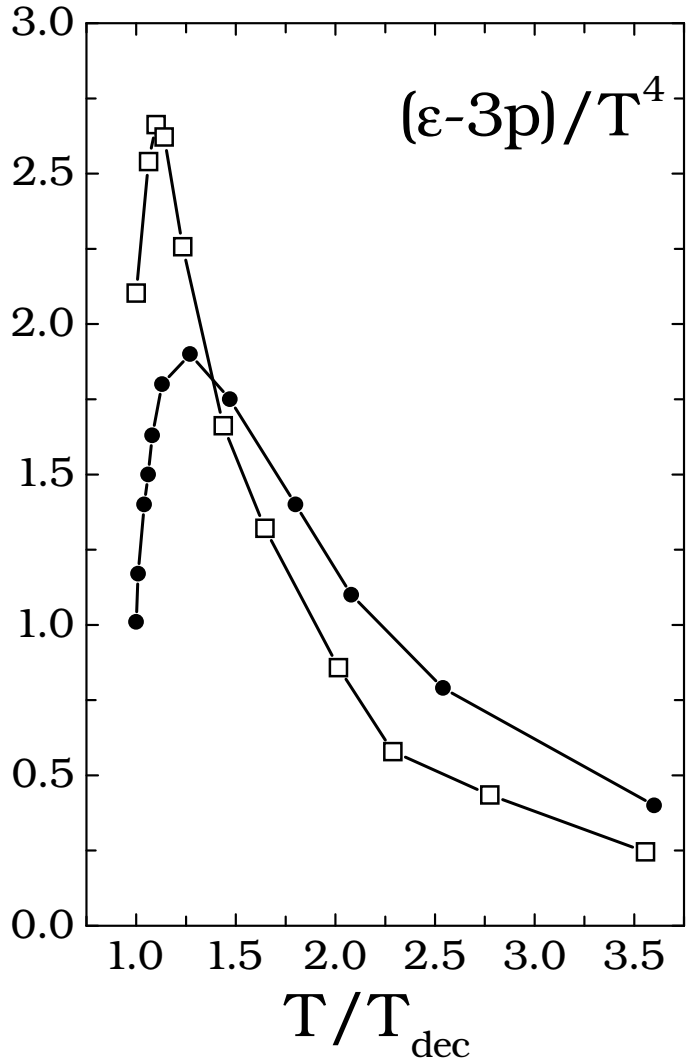}
\end{center}
\caption{The interaction measure $(\varepsilon-3p)/T^4$ of the
$SU(3)$ gluon system: circles are our data for the massless
gluons interacting via screened strings; squares show the lattice
results from Ref.\protect\cite{gboyd}. Our data are plotted for
the case $T_{\rm dec}=303\;MeV$.}
\label{fig2}
\end{figure}

In Fig. \ref{fig1} the quantities $\varepsilon/T^4$ and $3p/T^4$
are plotted as functions of the temperature for the system of
gluons with the spectrum Eq.~(\ref{sp1}) beyond the Boltzmann
approximation. As it is seen, Eq.~(\ref{tchem1}) indeed provides
a good estimation for the limiting temperature $T_{\rm chem}$,
which is now $303$ MeV. The deviation of the pressure and energy
curves from the ideal-gas value reflects essential attraction
even at $\approx 2T_c$. It is now interesting to clarify to what
extent our treatment of the in-medium string interactions agrees
with the nonperturbative lattice QCD. This can be understood with
the help of the special quantity $(\varepsilon-3p)/T^4$ that is
often called the interaction measure. This quantity is directly
related to remnant interactions that survive in the
high-temperature QCD phase because for the ideal massless quarks
and gluons $\varepsilon=3p$, as mentioned above. In Fig.\ref{fig2}
the interaction measure for the $SU(3)$ gluon plasma is shown. As is
seen, our treatment of the in-medium strings provides quite
reasonable results. The  quantity $(\varepsilon-3p)/T^4$ turns out
to  be very sensitive to the QCD interactions. Indeed, it is
still equal to zero even in the one-gluon-exchange approximation
provided the temperature dependence of the running coupling constant
is neglected. The agreement with the lattice calculations could be
even better if we chose $\gamma \approx 2/3$ like in
Ref.~\cite{lesush}. Thus, the interesting question arises what
additional arguments, being able to change the $\gamma$-value from
$1/3$ to $2/3$, should be taken into account for our picture of
the in-medium string screening. In this respect the spatial
correlations of color charges may be of importance~(see Sec. 3).

In the trans\-che\-mis\-try~\cite{BiLevZim} the reduced effective
value $\sigma = 0.5 \ GeV/fm = 0.1 \ GeV^2$ is used which results
in $T_{{\rm chem}}=185 \ MeV.$ It seems that the
trans\-che\-mis\-try model starts at  slightly lower temperature
where the chemical equilibrium for the quarks would not be
possible at all. On the other hand, this si\-mu\-lation begins
with a huge over\-sa\-tu\-ration of the quark number, so a later
reheating of the system brings the massive quark matter in an
over-critical state. Eventually, expansion and cooling leads to
dynamical hadronization at low temperatures ($T < T_{{\rm
chem}})$, where the quark component cannot be in chemical
equilibrium any more.

Now, returning to the Boltzmann statistics, for the critical number
density we have
\be
n_{\rm chem} = (\frac{2\lambda}{9e}\
\sigma )^{3/2} \ . \label{nchem}
\ee
Substituting this quantity into Eq.~(\ref{pdendep}) and taking
$\gamma = 1/3$,$\;A = (2/3)\sigma$, we get a negative pressure,
\be
p_{\rm chem} = - \frac{2 \lambda}{81 e} \sigma^2  \ ,
\label{pchem}
\ee
at the critical point. It means that the mechanical equilibrium
ceases at a somewhat higher temperature than the chemical one.

Equations similar to Eq.~(\ref{chem}) were considered by Boal,
Schach\-ter, Woloshin~\cite{olive} and by Mos\-ka\-len\-ko and
Khar\-ze\-ev~\cite{harz}, as well. How\-ever,
in\-ves\-ti\-ga\-tion of the quark plasma at zero baryon density,
presented in the latter paper, did not take into account the
important background term $\Phi$. As to the former one, only the
boundary for the high-temperature QCD phase, rather than the full
thermodynamics, was studied  without any reference to
the problem of thermodynamical consistency. This is why one of
the considered spectrum of unbound partons in this paper is not
consistent with Eqs.~(\ref{ctc1}) and (\ref{ctc1a}). In addition,
none of these papers uses the relevant classical approximation
providing analytical results like Eqs. (\ref{tchem}), (\ref{nchem}) and
(\ref{pchem}), which would have significantly simplified understanding.
At last, an advantage of our treatment is that it is based on
the elaborated model of the in-medium string formation that
enables us to derive the mean-field term~(\ref{meanf}) rather
than postulate it invoking the nearest-neighbor approximation
inspired by the analogy with the Ising model~(see the third paper
in Ref.~\cite{olive}).

\subsection{Fermions with string interaction}

Another interesting case is massless quarks at zero temperature. The
number density integral is given by
\be
n = \frac{d}{2\pi^2} \int \Theta (\mu - \frac{2}{3}\sigma n^{-1/3} -
k)\; k^2\, dk \ .
\label{T0quark}\ee
Here $d$ is the color, light flavor and spin degeneracy factor,
$\Theta(x)$ denotes the Heaviside step function. Expression
(\ref{T0quark}) can be readily rewritten as
$$
n =
\zeta^3 \ \left(\mu - \frac{2}{3}\sigma \ n^{-1/3}\right)^3
$$
with $\zeta=(d/6\pi^2)^{1/3}$. In the situation considered we have
one conserved charge: the baryon number with the density $n_b=n/3$.
Then, the chemical equilibrium is specified by the relation
$\mu=\mu_b/3$, where $\mu_b$ is the baryon chemical potential~(see
Sec. 2). The magnitude of $\mu_b$ is determined by the equation
\be
n_b=\frac{\zeta^3}{81} \left(\mu_b -
                   \frac{2}{3^{1/3}} \ \sigma \ n_b^{-1/3}\right)^3.
\ee
Using $z=\zeta n_b^{-1/3}/3^{3/4}$ we get
\be
\frac{1}{z} = \mu - \widetilde{\sigma}_1 \ z
\label{fstr}
\ee
with $\widetilde{\sigma}_1 = 6\sigma/\zeta.$
Equation~(\ref{fstr}) has a solution provided
\be
\mu_b \ge \mu_{{\rm chem}} = 2\sqrt{\widetilde{\sigma}_1},
\ee
{\em i.e.} the chemical potential (Fermi energy) is larger than
the minimum value of $(\widetilde{\sigma}_1 z+1/z)$. It means
that the Fermi energy of quarks should be larger than
$2\sqrt{\widetilde{\sigma}_1}/3$. A typical numerical value,
$\mu_{{\rm chem}} \approx 2.442$ GeV, can be found using $d=12$
and $\sigma = 0.18$ GeV$^2$.

\begin{figure}
\epsfxsize=120mm
\centerline{\epsfbox{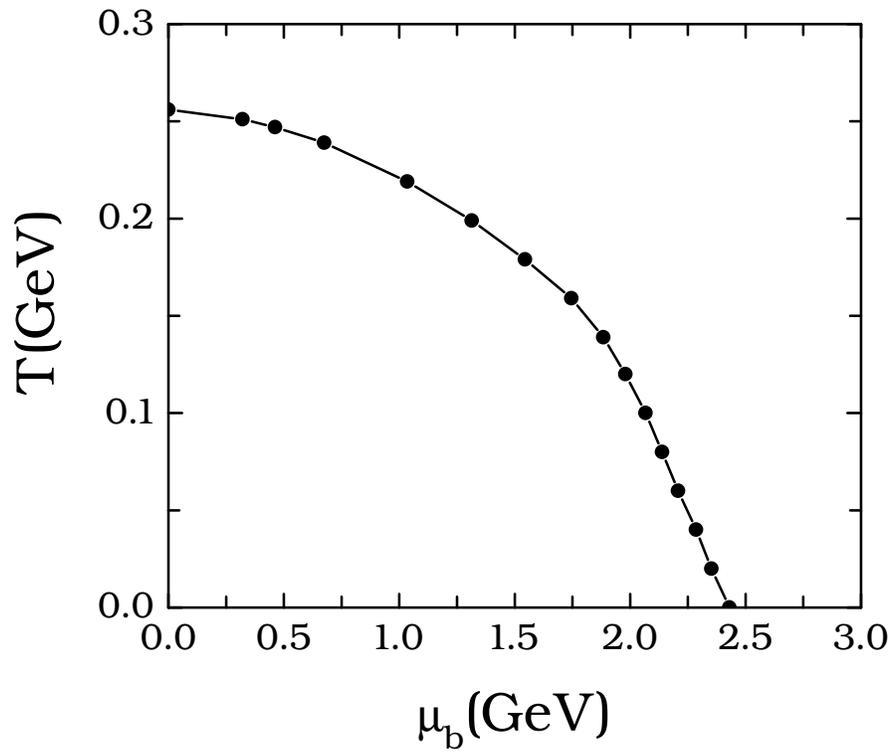}}
\caption{The critical curve on the temperature-chemical potential
 plane, below which chemical equilibrium ceases for a massless
 quark gas with string-like interaction.}
\label{fig3}
\end{figure}

At finite temperature we obtain a $T$- and $\mu_b$-dependent
consistency equation, which can be solved only numerically.
Fig.\ref{fig3} shows the resulting boundary in the
\hbox{$T$ -- $\mu_b$} plane.

\subsection{Chemical off-equilibrium in the classical approximation}

If an isolated system is out of chemical equilibrium and expands as
a perfect fluid, the relation
\be
dE+p \ dV = T \ dS + \sum_i\mu_i \ dN_i = 0 \
\ee
is fulfilled and the entropy production rate
\be
\dot{S} =-\sum_i\frac{\mu_i}{T}\, \dot {N_i} \
\ee is  either positive or zero.  This means that in the
one-component case for
quasiparticles with  positive $\mu$, the corresponding particle
number $N=Vn$ decreases while  it increases for negative values
of $\mu$.

\begin{figure}
\epsfxsize=100mm
\centerline{\epsfbox{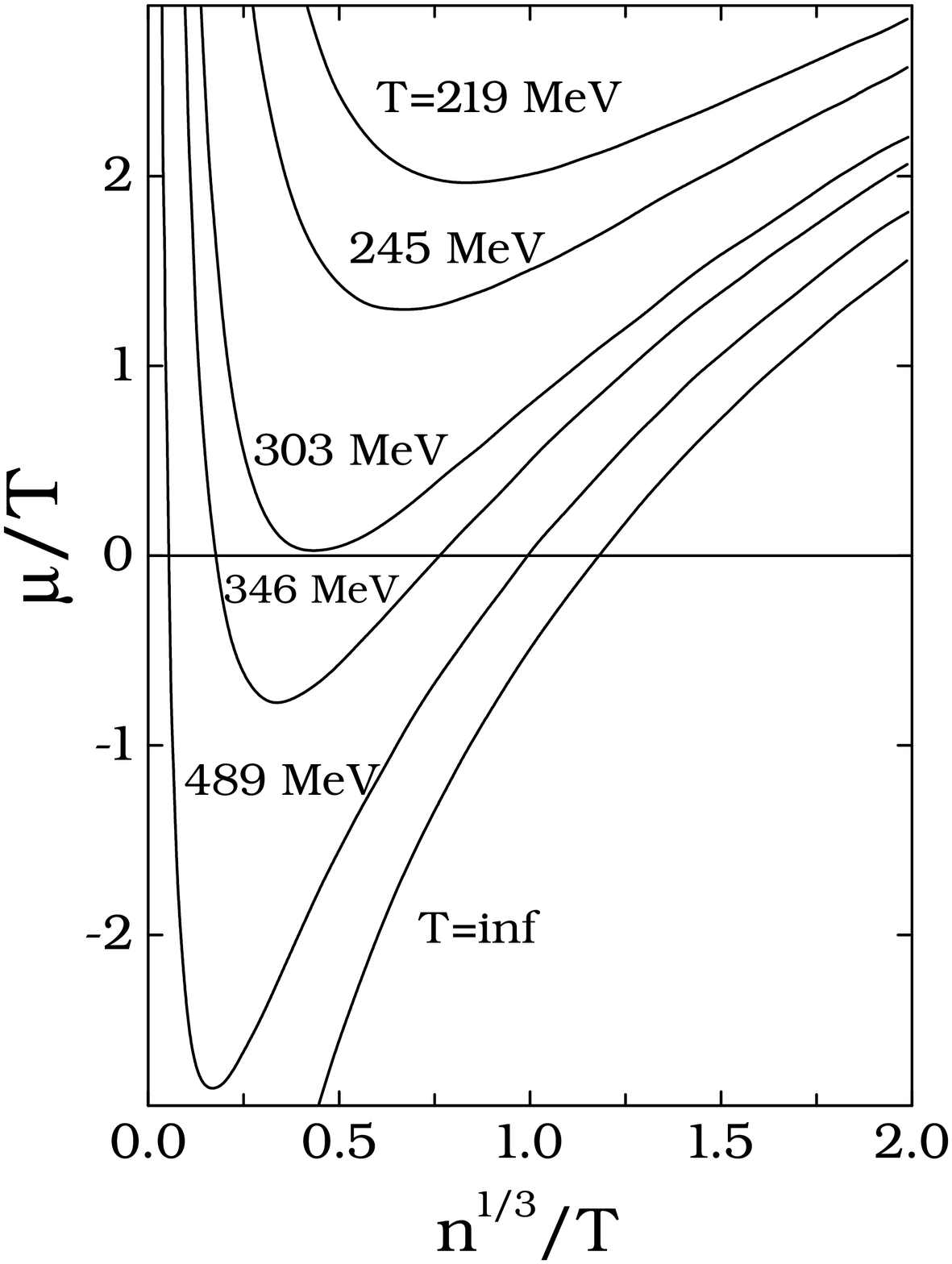}}
\caption{The scaled chemical potential $\mu/T$ as a function of
$n^{1/3}/T$ for massless Boltzmann gas with string-like interaction.
The chemical equilibrium condition is $\mu = 0$.}
\label{fig4}
\end{figure}

Fig.\ref{fig4} shows the off-equilibrium chemical potential
scaled with the temperature, $\mu/T,$ as a function of the scaled
density $n^{1/3}/T$ for a one-component, massless Boltzmann gas
made of particles with the $SU(3)$-gluon degrees of freedom and
interacting via strings. Chemical equilibrium corresponds to $\mu=0$,
which is not reachable below a certain temperature. Then the strings
pull the charges together never reaching a screened equilibrium
state: the chemical potential remains positive driving the density of
this component towards zero.

The situation is more complicated in a ma\-ny-com\-po\-nent mix\-ture
due to possible constituent exchange  between different species of
quasiparticles. In both cases the chemical equilibrium, corresponding
to $\mu_i =0$, is stable. In some special cases for non-ideal EoS it
may happen that the $\mu (n,T)$ curve for a constant $T$ (isoterm)
does not cross the $\mu =0$ line at all, {\em i.e.}  no chemical
equilibrium is possible and the system is driven towards a state with
either zero or infinite particle numbers. In a many-component system
it means that this particular component will dominate or vanish in
the mixture.

Another remark concerns with the chemical potential assigned to the
conserved charges ({\em e.g.} baryon number). This is {\em a physically
different} situation when the term $-\mu_b Q_b$~(see Sec. 2) is added
to the Hamiltonian, which is {\em not compensated} by its expectation
value in the background field. As a consequence, the chemical
equilibrium point (if any) is placed not at $\mu =0$ but at $\mu =\mu_B$.
This situation is quite accustom in nuclear physics.

\section{ Conclusions}

A useful representation  of the conditions of {\em thermodynamical
consistency} of quasiparticle description, Eqs.~(\ref{ctc1}) and
(\ref{ctc1a}), has been found. The advantage of this
representation is that it directly involves the effective
quasiparticle spectra, which results in important restrictions to
the form of these spectra. In particular, two essential findings
can be mentioned. If the interaction with surrounding matter is
taken into account by  introducing a mean field, the latter should
be either temperature independent (Example 1, Sec.2) or, when
in-medium effects are included into the Hamiltonian by means of
the effective mass (Example 2, Sec. 2), this mass should depend on
the temperature or/and the quasiparticle density exclusively
through the scalar density of quasiparticles. On the large market
of available phenomenological models these general restrictions in
majority of cases were used intuitively, but sometimes were
erroneously missed.

The structure of the thermodynamical potential, derived from the
medium-dependent Hamiltonian in a thermodynamically consistent
way, has been further detailized by implementing the string
picture for the interaction between generic constituents. With
this aim the elaborated mean-field model of the in-medium string
interactions has been developed. This model supports the
 use of the inverse power of the color charge
density in the color mean field (see Eq.~(\ref{meanf})) that was
introduced earlier by various authors assuming validity of the
nearest neighbour approximation~\cite{olive,ropke,harz}. Results
of our treatment of the in-medium strings are found to be in
reasonable agreement with the lattice data on QCD thermodynamics.
Further probing of these interactions can be an application of the
developed equation of state to (hydro)dynamical calculations
allowing direct comparison with observables. Some steps towards
this direction have been done recently by analyzing the excitation
function for nucleon directed flow~\cite{INNST} and the relation
of the 'softest point' of equation of state with chemical
freeze-out~\cite{TCNRS} in heavy ion collisions.

Along with other results, we would like to comment on the excluded
volume modification of the single particle energy. In our
treatment the conditions $C_T=0, \ C_1=0$ have no solution in this
case without some additional assumptions. A resolution of this
issue has been done in Ref.~\cite{Gorenstein} and in the first
paper of Ref.~\cite{syu}.
\vspace{1.0cm}

{\Large {\bf  Acknowledgments}}\\[3mm]
\noindent
This work  has been supported by the Hungarian National Fund for
Scientific Research OTKA (project No. T034269). A.A.S. and V.D.T.
are grateful to KFKI Research Institute and the heavy-ion group of
Prof. Zimanyi for the kind hospitality during their visits to
Hungary supported by the Fund for scientific collaborations
MTA-Dubna. V.D.T. was also partially supported by DFG (project 436
Rus 113/558/0) and  RFBR (grant 00-02-04012-NNIO).

Discussions with B.~M\"uller, D.~Blaschke, D.~Rischke, J.~Zim\'anyi
and P.~L\'evai are gratefully acknowledged.



\newpage
\begin{center}
{\bf  Figure Captions}
\end{center}

Fig.1. Normalized energy density and pressure of a massless
gluon gas with string-like interaction. Dashed line demonstrates
the Stephan-Boltzmann regime corresponding to the case of a gas
of noninteracting gluons. \\

Fig.2. The interaction measure $(\varepsilon-3p)/T^4$ of the
$SU(3)$ gluon system: circles are our data for the massless
gluons interacting via screened strings; squares show the lattice
results from Ref.\protect\cite{gboyd}. Our data are plotted for
the case $T_{\rm dec}=303\;MeV$. \\

Fig.3. The critical curve on the temperature-chemical potential
 plane, below which chemical equilibrium ceases for a massless
 quark gas with string-like interaction. \\

Fig.4. The scaled chemical potential $\mu/T$ as a function of
$n^{1/3}/T$ for massless Boltzmann gas with string-like interaction.
The chemical equilibrium condition is $\mu = 0$.\\

\newpage


\newpage
\begin{thebibliography}{99}

\bibitem{karsch} F.~Karsch, E.~Laermann, A.~Peikert, C.~Schmidt,
S.~Stickman,  Nucl. Phys. Proc. Suppl. {\bf 94},
411 (2001); F.~Karsch, E.~La\-er\-mann, A.~Peikert, Phys. Lett. B
{\bf 487}, 447 (2000); Nucl. Phys. B  {\bf 605}, 579 (2001).

\bibitem{MIT}  A.~Chodos, R.~L.~Jaffe, K.~Johnson, C.~B.~Thorn,
Phys. Rev. D {\bf 10}, 2599 (1974);
K.~Johnson, C.~B.~Thorn, Phys. Rev. D {\bf 13}, 1934 (1976);
K.~Johnson, Phys.Lett. B {\bf 78}, 259 (1978);
J.~F.~Donoghue, K.~Johnson, Phys. Rev. D {\bf 21}, 1975 (1980);
P.~Hasenfratz, J.~Kuti, Phys. Rept. {\bf 40}, 75 (1978);
B.~K.~Patra, C.~P.~Singh, Z. Phys. C {\bf 74}, 699 (1997);
R.~Hofmann, T.~Gutsche, M.~Schumann, R.~D.~Viollier,
Eur. Phys. J. {\bf C 16}, 677 (2000).

\bibitem{SLAC}  R.~Giles, Phys. Rev. D {\bf 13}, 1670 (1976).

\bibitem{CLOUDY}  G.~A.~Miller, A.~W.~Thomas, S.~The\-berge,
Phys. Lett. B {\bf 91}, 192 (1980);
S.~The\-berge, A.~W.~Thomas, G.~A.~Miller, Phys. Rev. D {\bf 22},
2838 (1980); {\it ibid} {\bf 23}, 2106 (1981);
A.~W.~Thomas, S.~The\-berge, G.~A.~Miller, Phys. Rev. D {\bf 24},
216 (1981);
S.~Lee, K.~J.~Kong, Phys. Lett. B {\bf 202}, 21 (1988);
G.~G.~Bu\-na\-tian, Yad. Fiz. {\bf 49}, 1071 (1989);{\em ibid} 1363
(1989)) (translated as Sov. J. Nucl. Phys. {\bf 49}, 664 (1989);
{\em ibid} 847 (1989));
Y.~Fu\-tami, S.~Aki\-yama, Prog. Theor. Phys. {\bf 84}, 377 (1990);
G.~A.~Mil\-ler, A.~W.~Tho\-mas, Phys. Rev. C {\bf 56}, 2329 (1997).

\bibitem{levai+heinz} P. Levai, U. Heinz, Phys. Rev. C
{\bf 57}, 1879 (1998).

\bibitem{Kampfer}  A.~Peshier, B.~K\"ampfer, G.~Soff, Phys. Rev. C
{\bf 61}, 045203, 2000;
B.~K\"ampfer, A.~Peshier, G.~Soff, J. Phys. G {\bf 27}, 535 (2001).

\bibitem{Hagedorn+Rafelski} J.~Rafelski, R.~Ha\-ge\-dorn,
QCD161:177:1980 (Symposium on Statistical Mechanics of Quarks
and Had\-rons, Bielefeld, Germany, August 24-31, 1980);
{\em ibid}, Phys. Lett.B {\bf 97}, 136 (1980); {\em ibid},
QCD161:W7:1978 (Workshop on Hadronic Matter at Extreme Energy
Density, Erice, Italy, Oct.13-21, 1978).

\bibitem{Gorenstein} J.~Cleymans, M.~I.~Gorenstein, J.~Stalnacke,
E.~Suhonen, Phys. Scr. {\bf 48}, 277 (1993); V.~K.~Tiwari,
N.~Prasad, C.~P.~Singh, Phys. Rev. C {\bf 58}, 439 (1998);
V.~K.~Tiwari, K.~K.~Singh, N.~Prasad, C.~P.~Singh, Nucl. Phys. A
{\bf 637}, 159 (1998); M.~I.~Gorenstein, A.~P.~Kostyuk,
Y.~D.~Krivenko, J. Phys. G {\bf 25}, L75 (1999); A.~Kostyuk,
M.~I.~Gorenstein, S.~Stoecker, W.~Greiner, Phys. Rev. C {\bf 63},
044901 (2001).

\bibitem{ton} E.~G.~Nikonov, V.~D.~Toneev, A.~A.~Shanenko,
Yad. Fiz. {\bf 62}, 1301 (1999) (translated as Phys. Atom. Nucl.  {\bf
  62}, 1226 (1999));
E.~Nikonov, A.~Shanenko and V.~Toneev, Heavy Ion Physics
{\bf 8}, 89 (1998).

\bibitem{Knoll}  H.~W.~Barz, B.~L.~Friman, J.~Knoll, H.~Schulz,
Nucl. Phys. A {\bf 484}, 661 (1988); {\em ibid}
Phys. Rev. D {\bf 40}, 157 (1989); {\em ibid}
Phys. Lett. B {\bf 242}, 328 (1990); {\em ibid}
Nucl. Phys. A {\bf 519}, 831 (1990).

\bibitem{BiLevZim} T.~S.~Bir\'o, P.~L\'evai, J.~Zim\'anyi,
J. Phys. G {\bf 25}, 1311 (1999); {\em ibid} Phys. Rev. C
{\bf 59}, 1574 (1999).

\bibitem{syu} A.~A.~Shanenko, E.~P.~Yukalova, V.~I.~Yukalov,
Physica A {\bf 197}, 629 (1993); {\em ibid} Yad. Fiz.{\bf 56}, 151
(1993) (translated as Phys. Atom. Nucl. {\bf 56}, 372 (1993)).

\bibitem{gor} M.~I.~Gorenstein, S.~N.~Yang, Phys. Rev. D {\bf
52}, 5206 (1995); {\em ibid}, J. Phys. G {\bf 21}, 1053 (1995).

\bibitem{Satz} T.~Celik, F.~Karsch, H.~Satz, Phys. Lett. B {\bf 97},
128 (1980); H.~Satz, Nucl. Phys. A {\bf 642}, 130 (1998);
S.~Fortunato, H.~Satz, Phys. Lett. B {\bf 475}, 311 (2000);
S.~Fortunato, F.~Karsch, P.~Petreczky, H.~Satz, Nucl. Phys. Proc.
Suppl. {\bf 94}, 204 (2001).

\bibitem{olive} K.~A.~Olive, Nucl. Phys. B {\bf 190}, 483
(1981); {\em ibid.} {\bf 198}, 461 (1982); D. H. Boal,
J.~Schachter, and R.~M.~Woloshin, Phys. Rev. D {\bf 26}, 3245
(1982).

\bibitem{ropke} D.~Blaschke, F.~Reinholz, G.~R\"opke, D.~Kremp,
Phys. Lett. B {\bf 151}, 439 (1985).

\bibitem{harz}  M.~Plumer,  S.~Raha and  R.M.~Weiner, Nucl.
Phys. A {\bf 418}, 549(1984);  V.~V.~Balashov,  I.~V.~Moskalenko
and  D.~E.~Kharzeev, Yad. Fyz. {\bf 47},
1740 (1988)) (translated as Sov. J. of Nucl. Phys. {\bf 47},
1103 (1988));
I.~V.~Moskalenko and  D.~E.~Kharzeev, Yad. Fiz.{\bf 48},
1122 (1988) (translated as Sov. J. of Nucl. Phys. {\bf 48},
713 (1988)).

\bibitem{kagiyam} S.~Kagiyama, A.~Minaka and A.~Nakamura,
Prog. Theor. Phys. {\bf 89}, 1227 (1993); {\em ibid} {\bf 95}, 793
(1996); {\em ibid}, hep-ph/0101036; {\em ibid}, hep-ph/0202108.

\bibitem{risch} D.~H.~Rischke, M.~I.~Go\-ren\-stein, H.~St\"ocker
and W.~Grei\-ner, Z. Phys. C {\bf 51}, 485 (1991); Q.~R.~Zhang, Z.
Phys. A {\bf 351}, 89 (1995).

\bibitem{prasad} N.~Prasad, K.~K.~Singh, and C.~P.~Singh, Phys. Rev.
C {\bf 62}, 037903 (2000).

\bibitem{bog} N. N. Bogoliubov, N.~N.~Bogoliubov, Jr., {\it
Introduction to Quantum Statistical Mechanics} (World Scientific,
Singapore,1982), see Secs.~3.4.1 and 3.4.2.


\bibitem{heinz}  U.~Heinz, W.~Greiner, W.~Scheild, J. Phys. G
{\bf 5}, 1383 (1979); U.~Heinz, P.~R.~Subramanian,
H.~St\"ocker and W.~Greiner, J. Phys. G. {\bf 12}, 1237
(1986).

\bibitem{kap} J.~I.~Kapusta, K.~A.~Olive,  Nucl. Phys. A {\bf 408},
478 (1983).

\bibitem{zim} J.~Zim\'anyi, B.~Luk\'acs, P.~L\'evai, J.~P.~Bon\-dorf,
N.~L.~Ba\-l\'azs, Nucl. Phys. A {\bf 484}, 647 (1988).

\bibitem{anchish} D.~V.~Anchishkin, Sov. Phys. JETP {\bf 75},
195 (1992); D.~Anchishkin and E.~Suhonen, Nucl. Phys. A {\bf 586},
734 (1995); V.~K.~Tiwary, N.~Prasad and C.~P.~Singh, Phys. Rev. C
{\bf 58}, 439 (1998).

\bibitem{walecka} J. D. Walecka, Ann. Phys.(N.Y.) {\bf 83},
491 (1974); {\em ibid}, Phys. Lett. B {\bf 59}, 109 (1975).

\bibitem{golov} V.~Goloviznin, H.~Satz, Z. Phys. C {\bf 57},
671 (1993).

\bibitem{rischke} D.~H.~Rischke, Nucl. Phys. A {\bf 583},
663 (1995).

\bibitem{pin} Bi~Pin-zhen, Shi~Yao-ming, Z. Phys. C. {\bf 75},
735 (1997).

\bibitem{shan} A.~A.~Shanenko, Phys. Rev. E {\bf 54}, 4420 (1996).

\bibitem{cleymans} E.~Suhonen, J.~Cleymans, J.~Stalnacke,
Nucl. Phys. Proc. Suppl. B {\bf 24}, 260 (1991);
J.~Cleymans, M.~Marais, E.~Suhonen, Phys. Rev. C {\bf 56}, 2747 (1997);
A.~Keranen, J.~Cleymans, E.~Suhonen, J. Phys. G {\bf 25}, 275, (1999).

\bibitem{gboyd} G.~Boyd et al., Nucl. Phys. B {\bf 469}, 419 (1996).

\bibitem{lesush} V.~D.~Toneev, E.~G.~Nikonov, and A.~A.~Shanenko,
in {\it Nuclear Matter in Different Phases and Transitions}, eds.
J.-P.~Blaizot, X.~Campi, and M.~Ploszajczak (Kluwer A. P., 1999)
p.309; Preprint GSI-98-30, Darmstadt, 1998.

\bibitem{INNST} Yu.~Ivanov, N.~Nikonov, W.~Norenberg, A.~Shanenko
and V.~Toneev, Heavy Ion Phys. {\bf 15}, 117 (2002).

\bibitem{TCNRS}V.~Toneev, J.~Cleymans, E.~Nikonov, K.~Redlich,
A.~Shanenko, J. Phys. G {\bf 27}, 827 (2001).

\end{thebibliography}
\end{document}